\documentclass[prl,twocolumn,showpacs,preprintnumbers,amsmath,amssymb]{revtex4}

\usepackage{graphicx}%Include Figures from File (\includegraphics) 

\begin{document}

%\preprint{}

\title{The influence of optical molasses in loading a shallow optical trap}

\author{Mathew S. Hamilton, Anthony R. Gorges, and Jacob L. Roberts}

\affiliation{Department of Physics, Colorado State University, Fort Collins, Colorado 80523}

\date{\today}

\begin{abstract}
We have examined loading of $^{85}$Rb atoms into a shallow Far-Off-Resonance Trap (FORT) from an optical molasses and compared it to loading from a Magneto-Optical Trap (MOT).
We found that substantially more atoms could be loaded into the FORT via an optical molasses as compared to loading from the MOT alone.
To determine why this was the case, we measured the rate of atoms loaded into the FORT and the losses from the FORT during the loading process.
For both MOT and molasses loading, we examined atom load rate and losses over a range of detunings as well as hyperfine pump powers.
We found that the losses induced during MOT loading were essentially the same as the losses induced during molasses loading at the same MOT/molasses detuning.
In contrast, load rate of the molasses was higher than that of a MOT at a given detuning.
This caused the optical molasses to be able to load more atoms than the MOT.
Optimization of FORT loading form an optical molasses improved the number of atoms we could trap by a factor of two over that of optimal loading from a MOT.

\end{abstract}

\pacs{blank}

\maketitle

%INTRODUCTION-------------------------------------------------------------------------------------------
Far-Off Resonance Traps (FORTs)\cite{Miller1993} have proven to be a valuable tool for ultracold experimentation.
FORTs have the advantage of being able to confine atoms in any magnetic sublevel for long periods of time without inducing heating from rescattered photons\cite{Miller1993}.  
For example, the creation of both Bose-Einstein Condensates (BECs)\cite{Barrett2001,Weber2003,Takasu2003,Cennini2003,Griesmaier2006,Dumke2006,Gericke2007,Gross2008,Beaufils2008} and quantum degenerate Fermi gases\cite{Granade2002,Fukuhara2007} without the use of a magnetic trapping apparatus can be accomplished using a FORT.
The ability of FORTs to trap any magnetic sublevel have enabled experiments using Feshbach resonances\cite{Loftus2002,OHara2002,Jochim2002,Marte2002,Regal2004} involving non-magnetically trappable states and BEC formation of Cs\cite{Vuletic1999,Yin2001,Weber2003,Cennini2006}.
FORTs even allow the capture and confinement of molecules\cite{Takekoshi1998,Jochim2003,Strecker2003,Zirbel2008,Ospelkaus2008}.

Loading as many atoms as possible into the FORT is one of the main considerations of most experiments involving FORTs.
For example, when cooling to degeneracy the starting number of atoms trapped is important.
Since it is common for a FORT to be loaded from a Magneto-Optical Trap (MOT), the physics of the loading of a FORT from a MOT has been the subject of many studies.
With the atom number requirement in mind, MOT loading into the FORT has been studied as a function of: MOT detuning \cite{Kuppens2000}, trap power\cite{Wu2006, Prime2007}, trap detuning\cite{Prime2007}, whether the FORT is pulsed or continuous-wave\cite{Shiddiq2008}, the ellipticity of the FORT\cite{Prime2007}, trap depth\cite{OHara2001},  hyperfine pump power\cite{Kuppens2000, Yang2007}, and trap geometry\cite{Ahmadi2005}.
Of particular note, in reference \cite{Kuppens2000} it was found that the number of atoms loaded into the FORT was determined by the balance between two competing processes: the load rate of atoms into the FORT from the MOT, and light-assisted collisional loss.
In other work, it was found that having as large a FORT volume as possible was helpful in trapping more atoms\cite{Ahmadi2005}.  
While other groups have used an optical molasses stage during the loading process to improve the number of atoms trapped in the FORT\cite{Wu2006, Lezec2006, Muller2007, Griffin2006, Newell2003, Dumke2006}, the details of how this improves the FORT loading have not received the same attention as the loading from a MOT.

In this paper we examine loading $^{85}$Rb into a shallow FORT from an optical molasses and compare it to loading from a MOT.  
In terms of experimental parameters, the only difference between the MOT and optical molasses loading conditions is the presence of the magnetic field from the anti-Helmholtz coils used to initially confine the atoms in the MOT: during ``MOT loading'' this field is on and during ``optical molasses loading'' (sometimes we will refer to this as simply ``molasses loading'') it is off.
Somewhat strikingly, we find radically different loading behavior as a function of whether or not the anti-Helmholtz field is on or off.  
The reason we are studying loading into a shallow, as opposed to deep, FORT is that a shallow FORT has a larger volume, all other things being equal, which allows more atoms to be trapped by the FORT.

%EXPERIMENTAL THEORY------------------------------------------------------------------------------------
In this work we studied FORT loading using Rb atoms, and as in previously reported work\cite{Kuppens2000}, the number of atoms trapped in the FORT is dependent on two competing processes: the rate of atoms loaded into FORT and collisional losses.  
The atom load rate is determined by the temperature of the atoms, the number of atoms which enter the load volume, and how effectively these atoms are cooled so that they may become trapped by the optical potential.  
The losses are primarily induced by light-assisted collisions.  
The number of atoms loaded into the FORT is characterized using the rate equation\cite{Kuppens2000}:
\begin{eqnarray}
\frac{dN}{dt}=R(t) - \beta '<n>N 
\label{rateEq}
\end{eqnarray}
where N is the number of atoms in the FORT, $<$n$>$ is the average density of atoms, R(t) is the load rate of atoms into the trap, and $\beta$' is an effective two-body loss coefficient.  
Single-body losses contribute much less on the timescales we use and can thus be ignored in our treatment.  
The load rate, R(t), has an observable time-dependence.  
The effective loss coefficient, $\beta$',  was found to exhibit little to no variation during the course of loading, and so we treat it as a constant in our analysis.

As in our studies, we were most concerned with the maximum number of atoms that could be loaded into the FORT.  
Thus, we were most interested in the load rate (R(t)) at the time when the peak number of atoms were loaded into the FORT.  
Using Equation (\ref{rateEq}) when the number of atoms loaded is at its maximum, it is possible to formulate a simple relationship between the peak number of atoms (N$_{peak}$), the load rate at the peak load time (R$_{peak}$), the trap volume V (where N = $<$n$>$V), and $\beta$':
\begin{eqnarray}
\beta '=\frac{R_{peak}V}{N_{peak}^{2}} 
\label{peakEq}
\end{eqnarray}
Because we are ultimately interested in maximizing the number of atoms loaded, N$_{peak}$, it is convenient to characterize a set of load conditions by the effective loss coefficient, $\beta$', and the load rate at the peak atom number, R$_{peak}$.
By characterizing the loading for a set of parameters with R$_{peak}$ rather than the varying R(t), it is easier to compare different experimental conditions to one another in a succinct manner.  
Our measurements consisted of values of R$_{peak}$ and $\beta$' which were compared for the optical molasses vs. MOT loading over a range of MOT or molasses detunings and hyperfine pump (HFP) beam powers.  

%EXPERIMENTAL SETUP-------------------------------------------------------------------------------------
The experiment for MOT loading of the FORT begins with a cloud of cold $^{85}$Rb atoms prepared in a MOT using standard techniques\cite{Raab1987}.
The FORT, a 30 W beam with a trap depth of 120 $\mu$K produced from a CO$_{2}$ laser, is turned on and off via an Acousto-Optical Modulator (AOM) and is overlapped with the MOT region.
Atoms from a thermal Rb vapor are collected in the MOT until 2x10$^{8}$ atoms have accumulated.
The laser beams of the MOT are retroreflected and detuned to 12 MHz to the red of the cycling transition of $^{85}$Rb during loading of the MOT.
The MOT has an average peak intensity of 2.5 W/cm$^{2}$ in each of the six trapping beams.
We load into the FORT from the MOT using two Compressed MOT (CMOT) stages.
The first CMOT stage increases the MOT laser beam detunings to 20 MHz.
The HFP power is also significantly reduced to a value which is varied as an experimental parameter between 1.2 and 20 $\mu$W/cm$^{2}$.
This CMOT stage lasts 13 ms.
We found that the load was not very sensitive to changes of detuning and duration of this stage; however, performance was better with the preliminary CMOT stage than without.
The final CMOT stage maintains the same HFP power as the previous, but the detuning and duration are varied as experimental parameters.
We refer to the detuning of this stage as the ``MOT detuning''.
During this stage, the FORT is turned on via the AOM, and atoms are loaded into the FORT (this is where what we call ``MOT loading'' occurs).
The turn on time and duration of the FORT are changed depending on the measurement being taken.
At the completion of the loading process, first the HFP laser, and then the MOT beams and anti-Helmholtz coils are turned off to put all atoms into the lower F=2 hyperfine state.
Loaded atoms are left in the optical trap 100 ms after the loading stage is complete, allowing atoms in the MOT which were not loaded into the FORT to fall away from the imaging region.
The atoms are then released from the FORT by turning off the beam via the AOM and allowed to expand for 5 ms prior to being imaged with a CCD camera using absorption imaging.
The atom number loaded into the FORT is determined from the resulting image.

\begin{figure}
%\vspace{68mm}
\includegraphics{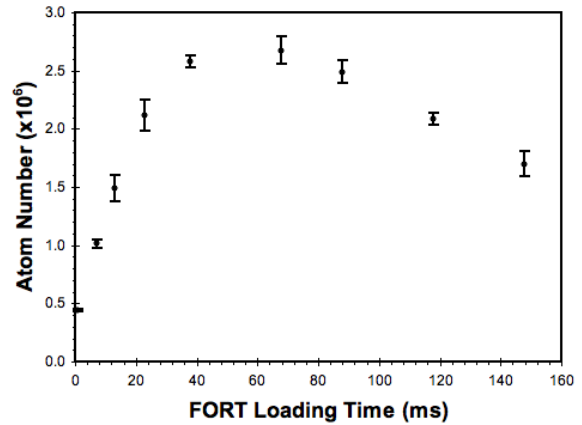}
\caption{\label{typical} Typical behavior of the number of atoms loaded into the FORT from the MOT as a function of loading time.  This particular evolution is the result of the final CMOT stage detuned to 33 MHz and a HFP power of 5 $\mu$W/cm$^{2}$.  In our analysis we will be primarily concerned with behavior at the peak atom number, which occurs in this case when atoms have been loaded from the final CMOT stage for 59 ms.}
\end{figure}

To extract R$_{peak}$ and $\beta$' for a given HFP power and MOT detuning, a series of different measurements are performed.
The first measurement taken is obtained by turning on the FORT beam at the start of the final CMOT stage.  
The CMOT and FORT remain on together for a variable amount of time, which we scan to obtain the number of atoms loaded into the FORT as a function of load time.
Typical behavior of this ``load evolution'' is depicted in Fig.~\ref{typical}.  
From the load evolution curve, we can extract the peak number of atoms, as well as the time at which the peak occurs.  

Once the peak number and time are known, we then take additional data to measure the loading rate, R(t) at the time corresponding to the maximum number of atoms loaded into the FORT (e.g. 59 ms in the case of Fig.~\ref{typical}).  
This is done by delaying the FORT turn on time to correspond to the time of maximum number of atoms in the load evolution.  
The number of atoms loaded into the FORT is measured for a range of short (2-6 ms) durations of MOT loading.  
Since the number of atoms in the FORT remains relatively small, behavior of dN/dt in Equation (\ref{rateEq}) is dominated by R(t) rather than the loss term.  
By combining these short duration measurements with the load evolution data, Eqns. (\ref{rateEq}) and (\ref{peakEq}) can be solved simultaneously to produce values for R$_{peak}$ and $\beta$'. 
This process is repeated for each set of experimental conditions (in the case of MOT loading, HFP power and detuning of the second CMOT stage).

\begin{figure}
%\vspace{53mm}
\includegraphics{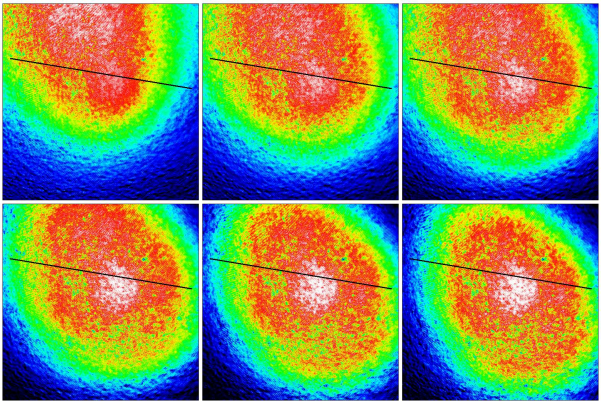}
\caption{\label{evo} (color) MOT images over a span of 73 ms.  The black line indicates the approximate position of the FORT.  As the MOT moves through space, its overlap with the FORT changes.  Optimal loading occurs when the position of the atom cloud has maximal overlap with the FORT during the loading stage.}
\end{figure}

Once $\beta$' is known, it is possible to measure R(t) at other times during the load evolution.
We observe that at the start of the CMOT stage R(t) increases for several milliseconds before subsequently decreasing.
This is due to a change in the position of the MOT with respect to the FORT during the CMOT stages.
Upon examination of the CMOT, we observe that it moves through space as time progresses (Fig.~\ref{evo}).
We speculate imperfections in the beam balance of our MOT/molasses lasers are responsible for the movement.
The exact form of the movement and final overlap depends on the precise detuning of the MOT/molasses lasers, as well as the HFP power.

When examining optical molasses loading, we follow the same basic procedure as we do when examining MOT loading.
The molasses loading experiment differs in several ways.
Most importantly, an optical molasses stage of adjustable duration appends the procedure for loading the FORT.
During this stage, HFP power remains the same, but the anti-Helmholtz coils are turned off, and the molasses laser (formerly the MOT laser) is further detuned an adjustable amount which we call the ``molasses detuning''.
The FORT beam is no longer turned on during the final CMOT stage, but instead during the following molasses stage.
The precise FORT turn on time and duration is again determined by the measurement taken.
During optical molasses loading measurements, the final CMOT stage is set to a fixed detuning of 33 MHz.
The final CMOT stage helps prepare the atom cloud location and density for loading into the FORT from the optical molasses.
The duration of the final CMOT stage becomes an additional variable in the molasses loading measurement.

To characterize the optical molasses loading for a chosen molasses detuning and HFP power, a series of measurements slightly different than the MOT loading measurements are taken.
For each set of experimental conditions, the final CMOT stage duration is optimized.
There are two factors which both depend on laser detuning and HFP power for this optimization: the motion and overlap of the CMOT with the FORT, as discussed earlier; and the decay of atom number held in the CMOT as a function of time.
To determine the optimal timing for the final CMOT stage, we measured out a ``load evolution'' curve which  is slightly different from the one shown in Fig.~\ref{typical}.  
We varied the duration of the final CMOT stage, but in addition to the CMOT, we added an optical molasses stage with a fixed length of time.  
In these measurements, the FORT is turned on at the start of the final CMOT stage, and turned off 100 ms after the completion of the molasses stage.  
The duration of the optical molasses stage is chosen iteratively so that in the final set of measurements this duration corresponds to the time during optical molasses loading at which the peak number of atoms are obtained in the FORT.

Once the final CMOT stage duration is set, we then proceed to take measurements in the same manner as with MOT loading, starting by turning the FORT on at the start of the molasses stage and taking data as we vary the time in which both the FORT and molasses stage are active to obtain the molasses load evolution.
The molasses loading evolution curve has a similar shape as the MOT loading evolution (Fig.~\ref{typical}).
The peak atom number of this evolution again gives the time to take the short duration loading data to determine R$_{peak}$ as well as the peak number of atoms loaded into the FORT.
Once the time of the load evolution peak is known, R$_{peak}$ and $\beta$' are found by setting the FORT turn on time to the time of the peak and following the same procedure as we did with MOT loading to solve Equations (\ref{rateEq}) and (\ref{peakEq}).

\begin{figure}
%\vspace{86mm}
\includegraphics{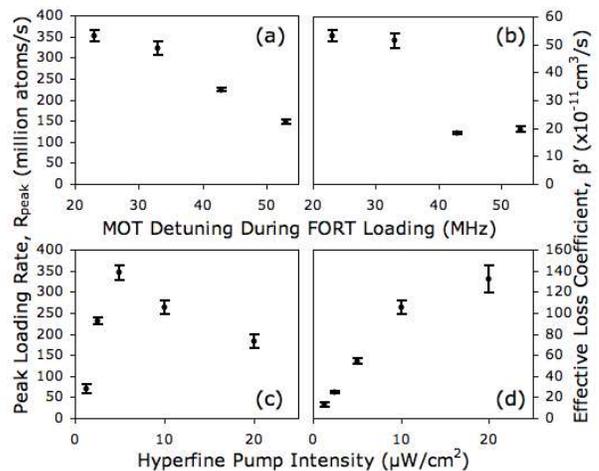}
\caption{\label{MOT} A set of results for the MOT loading experiment.  Behavior of (a) R$_{peak}$ versus MOT detuning, and (b) $\beta$' versus MOT detuning.  Detunings are in MHz to the red of the cycling transition at a set HFP intensity of 5 $\mu$W/cm$^{2}$.  (c) R$_{peak}$ as a function of HFP intensity, and (d) $\beta$' measured over the same range of HFP intensities.  The observation of the unexpectedly rapid change in $\beta$' between 33 and 43 MHz observed in (b) is consistent with other sets of data which also support that losses decrease with detuning.  The HFP intensities are in units of $\mu$W/cm$^{2}$, with a fixed MOT detuning of 33 MHz.  Load rates are in units of million atoms per second, loss rates are in units of cm$^{3}$s$^{-1}$.  Error bars depict statistical uncertainties.}
\end{figure}

%RESULTS------------------------------------------------------------------------------------------------
We collected load and loss rate data for both MOT and molasses loading over a wide range of detunings, as well as HFP powers.  
Values for both R$_{peak}$ and $\beta$' were plotted over the range of parameters to observe the trends in loading efficiency.
Figure~\ref{MOT} depicts the data for a particular set of conditions of loading from the MOT.
The left two plots show trends in the peak load rate, R$_{peak}$, while the right plots depict trends in the loss coefficient, $\beta$'.
In the top two plots we see the effect of MOT detuning at a fixed HFP power.
The lower two plots, the effect of HFP power on the load is shown for fixed MOT laser detuning.
We find that as detuning is increased, the MOT load rate decreases.
This decrease is expected since the increased detuning will mean that the atoms will spend more time in the lower hyperfine state (due to an increased likelihood of off-resonant transitions) causing the cooling scattering rate to decrease, limiting the effectiveness with which the atoms can be slowed by the MOT laser and subsequently trapped by the FORT.
As detuning is increased, we find that $\beta$' also decreases.
$\beta$' should decrease with further detuning for two reasons: first, the reduction of upper hyperfine ground state population reduces the light-assisted collision rate caused by the trapping light, and second, the light-assisted cross section itself can be affected by the detuning of the trapping light.

Increasing HFP power at a fixed MOT detuning at first causes an increase in the load rate, but eventually, the rate turns over and decreases with additional HFP power.
We speculate that this ultimate decrease in load rate with increasing HFP power is due to a drop off in cooling efficiency in the dense cloud due to rescattering and relatively larger average light forces which disrupt the MOT/FORT overlap.
Increasing HFP power increases the losses induced in the MOT.
This behavior is understood by the increase in upper hyperfine ground state population as a result of increased HFP power.
The increase in upper hyperfine ground state population increases the light-assisted collision rate and thus, the effective loss coefficient, $\beta$'.

\begin{figure}
%\vspace{68mm}
\includegraphics{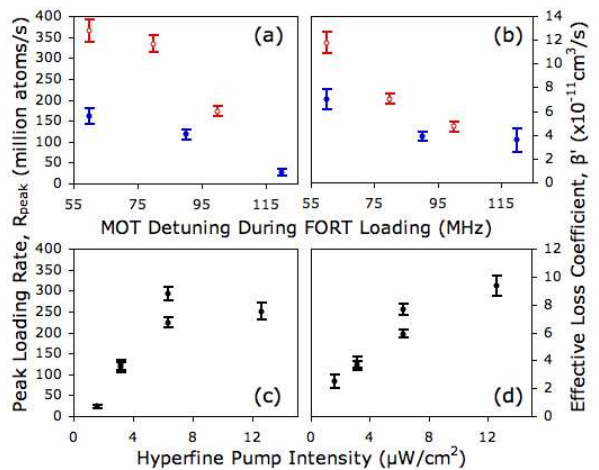}
\caption{\label{MOL} (color) Behavior of (a) R$_{peak}$, and (b) $\beta$' as a function of optical molasses detuning in MHz.  The data depicted as open (red) circles has a HFP intensity of 6.3 $\mu$W/cm$^{2}$ while the curve represented by filled (blue) circles is at an intensity of 3.2$\mu$W/cm$^{2}$.  The effect of HFP power on (c) R$_{peak}$, and (d) $\beta$' for a fixed molasses detuning of 90 MHz.  HFP power is in units of $\mu$W/cm$^{2}$.  Load rates are in units of million atoms per second, loss rates are in units of cm$^{3}$s$^{-1}$.  Error bars depict statistical uncertainties.}
\end{figure}

A set of loading behavior data for molasses loading is depicted in Fig.~\ref{MOL}.
The layout of Fig.~\ref{MOL} is the same as Fig.~\ref{MOT}, with the exception that there are two detuning curves shown: one with twice the HFP power of the other.
Although the behavior with detuning has the same general trends as that observed during MOT loading, we find that the range of detunings which result in effective optical molasses loading is higher in magnitude than the range of detunings where MOT loading is most efficient.
We find that the behavior of both R$_{peak}$ and $\beta$' scale with increasing molasses detuning in the same way as they do with increasing MOT detuning.
Both R$_{peak}$ and $\beta$' increase with increasing HFP power.

We note that over the course of our experiment, measured values of R$_{peak}$ and $\beta$' tended to vary slightly.
These changes in recorded data were observed on two time scales: a daily variation consistent with temperature fluctuations in the lab, and an overall drift observable over weeks or months attributed to the inevitable drift of the experimental apparatus.
We speculate that small changes in the alignment of our system can cause a significant effect in the magnetic sublevel distribution of the atom cloud.
This leads to changes in the effective optical pumping rates, affecting both the collisional losses as well as the effective cooling rate and thus the loading rate.
For this reason, the comparison data, such as that displayed, was taken in relatively short time spans.
The effect of alignment on the load rate and losses was checked explicitly, and is consistent with observations made during the experiment.

Outside of the detuning ranges of MOT/molasses detunings shown in Figures~\ref{MOT} and~\ref{MOL}, the performance of the loading of the optical trap is much poorer, inhibiting direct comparisons of MOT and molasses loading under otherwise the same experimental conditions.
Nevertheless, mild extrapolations of the data can be done to understand the differences in MOT and molasses loading.
The mechanism responsible for the change in behavior between MOT and optical molasses loading primarily manifests itself in the load rate.
Comparing the HFP power data between MOT and molasses loading, we see that the load rates are nearly the same between the two, even though the detunings used for each are very different (43 MHz for Fig.~\ref{MOT}(c) and 90 MHz for Fig.~\ref{MOL}(c)).
Given the observed decrease in R$_{peak}$ with increasing detuning for the MOT loading, this indicates that the molasses loading is far more effective at large detunings than for MOT loading.
For instance, a mild extrapolation of the data presented in Fig.~\ref{MOT}(a) and the open red points of Fig.~\ref{MOL}(a) which were taken with nearly the same HFP power, indicates that the loading rates at the peak atom number for the molasses are far greater than those of the MOT.

In contrast, the values for $\beta$' between the MOT and molasses loading cases are more consistent with $\beta$' being determined by only the HFP power and laser detuning and not whether or not the anti-Helmholtz coils are employed.  
Extrapolation of the data indicates that values of $\beta$' for both MOT and optical molasses loading are consistent with being part of a continuous behavior.
This can be seen in the unfilled red data points in Fig.~\ref{MOL}(b) which extrapolate to values of $\beta$' that are in the same range as those observed during MOT loading (Fig.~\ref{MOT}(b)).
The behavior of $\beta$' in response to HFP power is also the same between MOT and optical molasses loading.

The ultimate goal is to optimize the maximum number of atoms trapped in the FORT.  
Intuitively, this is accomplished by maximizing the load rate while minimizing the loss rate.  
However, both R$_{peak}$ and $\beta$' scale in a similar manner with both laser detuning and HFP power.
This causes the optimum to depend on the relative slopes of these dependencies, and can be analyzed by looking at the ratio R$_{peak}$/$\beta$' (which is related to the peak number of atoms, N$_{peak}$).
Using this figure of merit, our setup achieves optimal MOT loading at a detuning of 43 MHz, while the optical molasses optimizes at 80 MHz.
The load rate happens to be nearly the same value for both MOT and molasses at optimal loading.
The maximum number of atoms is then determined by the value of $\beta$' at each detuning.
Loading is maximized at nearly the same HFP power for both MOT and molasses loading.  
The maximum ratio of R$_{peak}$ over $\beta$' for our experimental apparatus is calculated to be greater for optical molasses loading than with MOT loading.
This calculation is consistent with an observed increase from about 2.5 million atoms when loading from a MOT alone, to over 5 million atoms when loading using an optical molasses stage.

The observed drifts in R$_{peak}$ and $\beta$' did not have a significant effect on the observed values of N$_{peak}$.
The reason behind this is that the drifts of both measured quantities changed in the same general way, keeping the ratio of R$_{peak}$ over $\beta$' roughly the same.
Furthermore, the optimal detunings maintain their values reasonably well despite the changes in R$_{peak}$ and $\beta$'.
This is only true when the alignment changes slightly; major misalignments clearly hinder loading efficiency and reduce the total number of atoms loaded into the FORT.

Thus far, we have only reported results where the FORT was loaded purely during the last CMOT stage or purely during the molasses stage.
We would expect that the last CMOT stage could be used as a ``boost'' to the molasses loading by turning on the FORT before the molasses and during the final CMOT stage.
This would allow the molasses to start loading with a non-zero atom number already in the optical trap, and result in a greater maximum number of atoms loaded into the FORT after the molasses stage.
We find that when the optical molasses is near optimum, a purely molasses loaded trap has nearly the same number of atoms as a trap loaded with both the final CMOT stage and the molasses stage.
In contrast, for non-optimal molasses conditions, the final CMOT and molasses stages share non-trivial contributions to total number loaded.
The reason for this is that R(t) decreases only slowly for the molasses loading.
Waiting to load additional atoms via the molasses instead of the CMOT thus does not affect the optimum significantly (i.e. by less than 5$\%$).
This implies that a well optimized molasses will essentially do all of the loading of atoms into the FORT.
Examination of non-optimal CMOT timings support this, since we find that the molasses stage can make up the losses of a non-optimal CMOT stage.
This also makes our system fairly resilient to significant change when it is near optimal conditions.

\begin{figure}[t]
%\vspace{51mm}
\includegraphics{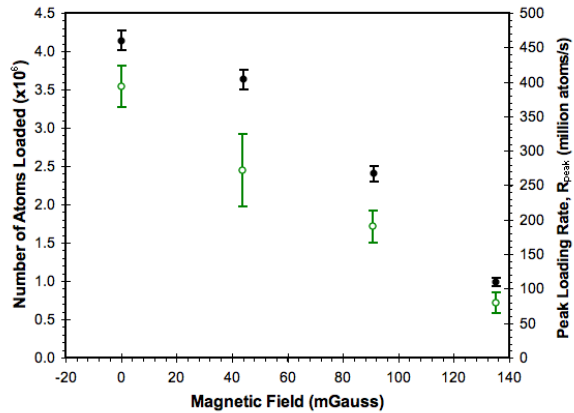}
\caption{\label{ExtF} (color) The effect of an applied uniform external magnetic field in milliGauss.  The left axis and filled black circles show the effect on atom number loaded into the FORT in units of million of atoms.  The right axis and open green circles show the effect on load rate into the FORT in units of million of atoms per second.  The reduction of atom number with external magnetic field is consistent with observed decreases of atom number loaded entirely from a MOT of similar magnetic field strength compared with the number of atoms loaded entirely from an optical molasses.  Because the magnetic field causes atoms to cool to non-zero velocity ($\sim$8 cm/s for 100 mG), atoms in the optical trap region are not cooled as effectively, causing the load rate to decrease.  Error bars depict statistical uncertainties.}
\end{figure}

The only difference between a MOT and optical molasses is the presence of the anti-Helmholtz magnetic field.
To investigate how the loading could depend on only the presence of the magnetic field, we applied a uniform field (rather than an anti-Helmholtz field) to the atom cloud while loading the FORT from an optical molasses.
We found that this applied magnetic field reduces the loading rate, which reduces the number of atoms loaded into the FORT (Fig.~\ref{ExtF}).
Checking the overlap of the MOT with the FORT revealed no significant change as a result of the applied field, indicating that the reduction is not position based.
The magnitude of the applied field required to significantly decrease the loading rate was on the order of 100 mG.
Using a loose approximation derived from the theory in reference \cite{vanderStraten1993}, we find that external fields of 100 mG will cause atoms to be cooled to velocities on the order of those found in the the optical molasses stage during loading.
Since the magnetic field gradient in the MOT along the direction of the FORT (one of the radial MOT directions) is about 3 G/cm in our apparatus, the 100 mG is equivalent to the field at a distance of 0.33 mm from the center of the anti-Helmholtz field in the FORT trapping region during the CMOT stage.
Because the axial extent of the atom cloud in the FORT has a rms size of 0.77 mm, we expect that cooling in the edges of the atom cloud far away from the zero of the magnetic field is less efficient, and causes a reduction of the load rate where the optical trap intersects this region.
A rough estimate of the reduction of cooling based on the geometry of our optical trap and the anti-Helmholtz field in our setup indicates that this effect is consistent with the observed reductions.
Examination of the atom cloud temperature in the MOT and molasses reveals that the anti-Helmholtz coils cause the temperature of the atoms to rise to about 30$\mu$K from a value less than 20$\mu$K.
The fact that the magnetic field causes atoms to cool to non-zero velocity is enough to account for the observed decline in the load rate.

%CONCLUSION---------------------------------------------------------------------------------------------
In summary, the loading of a FORT from an optical molasses loads more atoms than loading from a MOT.
This is due to the fact that the load rate for optical molasses loading is much higher than that of MOT loading.
The difference in the load rates cause molasses loading to optimize at a higher detuning than optimal MOT loading.
Because the losses decrease with higher detuning, the losses for optimal molasses loading are lower than the losses for optimal MOT loading.
These effects are significant enough that loading with a molasses improves the number of atoms we can load into our FORT by a factor of two.

This work is funded by the Air Force Office of Scientific Research, grant number FA9550-06-1-0190.

\bibliographystyle{apsrev.bst}

\end{document}